\documentclass[a4paper,twoside,reqno]{bjp}
\usepackage{graphicx}
\usepackage{cite}
\usepackage{amssymb,amsmath,amscd,amsthm}
\usepackage{times}

\usepackage[bookmarks=false]{hyperref}
\hypersetup{%
    colorlinks=true,        % false: boxed links; true: colored links
    linkcolor=blue,          % color of internal links (change box color with linkbordercolor)
    citecolor=blue,         % color of links to bibliography
    urlcolor=blue           % color of external links
    }

\usepackage{geometry}
\allowdisplaybreaks

\begin{document}

\title{Study of resonances in exotic drip-line nuclei by the use of super-symmetric quantum mechanics}%
% Insert the title

\runningheads{Hyperspherical three-body model calculation for the ground and ....}{Md. A. Khan et al.}
% In case the authors are more than three, put the name of the first author followed by the Latin `et al.'

\begin{start}{%
\author{Md. A. Khan*}{1},
% The second argument connects author(s) with addresses
\author{M. Hasan}{1},
% The second argument connects author(s) with addresses
\author{S. H. Mondal}{1},
\author{M. Alam}{1},
\author{T. Surungan}{2}

\address{Department of Physics, Aliah University, \\ IIA/27, Newtown, Kolkata-700160, India; (*Corresponding author; E-mail: drakhan.phys@aliah.ac.in)}{1}
\address{Department of Physics, Hasanuddin University, \\Makassar-90245, Indonesia}{2}
%\address{Affiliation of the author; the second argument connects the address with author}{2}

\received{Day Month Year (Insert date of submission)}
% Insert date of submission
}

\begin{Abstract}
In this paper, a nice theoretical scheme is presented to investigate resonant and bound states in weakly bound nuclear systems by the use of isospectral potentials together with hyperspherical harmonics expansion. In this scheme, a new potential is constructed which is strictly isospectral with the original shallow potential and has properties that are desirable to calculate resonances more accurately and elegantly. The effectiveness of the method has been checked in terms of its application to the $0^+$ resonances in the neutron-rich isotopes $^{18,20}$C in the two-neutron plus core three-body cluster model. The results are in excellent agreement with the available data.
\end{Abstract}

\begin{KEY}
  Halo nuclei, resonance, isospectral potential.
\end{KEY}
\end{start}

%%%%%%%%%%%%%%%%%%%%%%%%%%%
\section{Introduction}
The discovery of halo nuclei (neutron-halo and proton-halo) in the neighborhood of the nuclear drip lines is one of the major achievements since the development of the radioactive ion-beam facilities. This exotic kind of nuclear structure is characterized by a dense stable core surrounded by a low-density envelope of long extension. This low-density envelope is supposed to be the consequence of quantum mechanical tunneling of the last nucleon(s) through a-wide and skinny barrier following an attractive shallow potential well. In halo nuclei, one seldom finds any excited bound state due to this typical nature of the potential. These nuclei utmost support one bound state at energies of the order of 1 MeV. One of the most exciting features of halo nuclei is the fact that they exhibit one or more resonance state(s) (which has high scientific significances) just above the binding threshold.

In the nuclear landscape, one may find some of the observed one-neutron halo nuclei-$^{17}$B, $^{19}$C; two-neutron halo nuclei- $^6$He, $^{11}$Li, $^{11, 14}$Be; one-proton halo nuclei- $^8$B, $^{26}$P, $^{20}$C; two-proton halo nuclei-$^{17}$Ne, $^{27}$S and also four-neutron halo nuclei-$^{14}$Be, $^{19}$B, etc \cite{tanihata-1985} \cite{schwab-1995} \cite{jensen-2000} \cite{tanaka-2005} \cite{canham-2008} \cite{tanaka-2010} \cite{kobayashi-2012} \cite{hwash-2017}. Due to low-density envelope surrounding the dense core, halo-nuclei have unusually large RMS matter radii (larger than the liquid-drop model prediction of $R_A\propto A^{1/3}$) as reported by Audi et al 2003 \cite{audi-2003}, Acharya et al 2013 \cite{acharya-2013}. And due to weaker effective interaction among the valence nucleons and the core, these nuclei exhibit sufficiently small two-nucleon separation energies (typically less than 1 MeV). Till now, much scientific research has been conducted to study the structure of this exciting nuclear species. Tanaka et al 2010 \cite{tanaka-2010} observed of a large reaction cross-section in the drip-line nucleus $^{22}$C, Kobayashi et al 2012 \cite{kobayashi-2012} conducted research on one- and two-neutron removal reactions using the most neutron-rich carbon isotopes, Gaudefroy et al 2012 \cite{gaudefroy-2012} carried a direct mass measurements of $^{19}$B, $^{22}$C, $^{29}$F, $^{31}$Ne, $^{34}$Na and some other light exotic nuclei. Togano et al, 2016 \cite{togano-2016} studied interaction cross-section of the two-neutron halo nucleus $^{22}$C. \\In the literature review, it is noted that for the investigation of the structure of the 2n-halo nuclei three main theoretical approaches are used- i) the microscopic model approach where the valence neutrons are supposed to move around the conglomerate of other nucleons (protons and neutrons) without having any stable core, S\"{a}\"{a}f et al \cite{saaf-2014}; ii) the three-body cluster model in which the valence nucleons are assumed to move around the structureless inert core, and iii) the microscopic cluster model in which the valence nucleons move around the deformed excited core as reported in \cite{saaf-2014} \cite{nesterov-2010}\cite{korennov-2004}. One may also refer to the renormalized zero-range three-body model of Souza et al 2016 \cite{souza-2016}. And for the computation of resonant states, several theoretical approaches are employed. Some of those include the positive energy solution of the Faddeev equation by Cobis et al 1997 \cite{cobis-1997}, complex coordinate rotation (CCR) by Csoto 1993 \cite{csoto-1993}, Ayoma et al \cite{ayoma-1995}, the analytic computation of bound state energies by Tanaka et al 1997 \cite{tanaka-1997}, the algebraic version of resonating group method (RGM) by Vasilesky et al 2001 \cite{vasilevsky-2001}, continuum-discretized coupled-channels (CDCC) method clubbed to the cluster-orbital shell model (COSM) by Ogata et al 2013 \cite{ogata-2013}, hyperspherical harmonics method (HHM) for scattering states by Danilin et al 1997 \cite{danilin-1997}, etc. In most of the theoretical approaches, Jacobi coordinates are used to derive the relative coordinates after the separation of the center of mass motion.\\One of the most challenging obstacles that people often encounter in the calculation of resonances in any weakly bound nucleus is the large degree of computational error. In our case, we shall overcome this major obstacle by adopting an elegant theoretical approach through interfacing the algebra of supersymmetric quantum mechanics with the algebra involved in the hyperspherical harmonics expansion method. In this scheme, one can handle the ground state as well as the resonant states on the same footing. The technique is based on the interesting fact that, for any arbitrarily given potential (say, {$U$}), one can construct a family of isospectral potentials ($\hat{U}$), in which the latter depends on an adjustable parameter ($\lambda$). And when the original potential has a significantly low and excessively wide barrier (poorly supporting the resonant state), $\lambda$ can be chosen judiciously to enhance the depth of the well together with the height of the barrier in $\hat{U}$. This enhanced well-barrier combination in $\hat{U}$ facilitates trapping of the particle of interest and this, in turn, facilitates the computation of resonant state more accurately at the same energy, as that in the case of the original potential {$U$}. This is possible because, {$U$} and $\hat{U}$ have {\bf strictly isospectral} nature.\\To validate the effectiveness of the scheme we will apply the proposed scheme to the first $0^+$ resonant states of the carbon isotopes $^{\rm A}$C, for mass number {\rm A} = 18 and 20 respectively. We chose the three-body (2n + $^{\rm A-2}$C) cluster model for both of the above isotopes, assuming that the outer core valence neutrons move around the inert core $^{A-2}$C. The lowest eigen potential derived for each of the three-body systems exhibits a shallow well following a skinny and excessively wide barrier. This skinny-wide barrier results in a large resonance width of the resonance. One can, in principle, find quasi-bound states in such a typical shallow potential, but that poses a difficult numerical task. For a finite height of the barrier, a particle can temporarily be trapped in the shallow well when its energy is close to the resonance energy. But, there is a finite possibility that the particle may creep in and tunnel out through the barrier. Thus, a more accurate calculation of resonance energy is easily masked by the large width of resonance resulting from a large tunneling probability due to a low barrier height. Hence, a straightforward calculation of the resonance energies of such systems fails to yield accurate results.
 
For the ground state energy and wavefunctions, we adopt the hyperspherical harmonics expansion method (HHEM) described by Fabre et al 1982 \cite{fabre-1982} to solve the three-body Schr\"{o}dinger equation in relative coordinates. In HHEM, three-body relative wavefunction is expanded in a complete set of hyperspherical harmonics. The substitution of the wavefunction in the Schr\"{o}dinger equation and use of orthonormality of HH gives rise to an infinite set of coupled differential equations (CDE). The method is an essentially exact one, involving no other approximation except an eventual truncation of the expansion basis subject to the desired precision in the energy and the configuration of the available computer. However, the hyperspherical convergence theorem prescribed by Schneider 1972 \cite{schneider-1972} permits extrapolation of the data computed for the finite size of the expansion basis, to estimate those for an even larger expansion basis. The fact that convergence in HH expansion is significantly slow and one needs a large number of CDE's to solve for the achievement of desired precision pauses practical a limitation to the method. But we used the hyperspherical adiabatic approximation (HAA) of Ballot et al 1982 \cite{ballot-1982} to construct a single differential equation (SDE) to be solved for the lowest eigen potential, ${U}_0(\rho)$) to get the ground state energy $E_0$ and the corresponding wavefunction $\psi_0(\rho)$ following method of Das et al 1982 \cite{das-1982}. 

We next derive the isospectral potential $\hat{U}(\lambda,\rho)$ using the ground state energy ad wavefunctions using algebra of the SSQM reported by Cooper et al 1995 \cite{cooper-1995}, Khare et al 1989 \cite{khare-1989}, and Nieto 1984 \cite{nieto-1984}. Finally, we solve the SDE for $\hat{U}(\lambda,\rho)$ for some positive energies for wavefunctions $\hat{\psi}(\lambda, \rho)$. We then compute the probability of trapping of the particle(s) in terms of the norm of the positive energy wavefunctions within the enhanced well-barrier region. A plot of the trapping probability as a function of energy shows a sharp peak at the resonance energy. The actual width of resonance can be obtained by back-transforming the wave function $\hat{\psi}(\lambda, \rho)$ corresponding to $\hat{U}(\lambda,\rho)$ to $\psi(\rho)$ of $U(\rho)$. Hence, the major goals of this work are three-fold: i) to reproduce the ground state spectra of the known systems, ii) to construct the isospectral potential using ground state energy and wavefunction, and iii) to obtain resonance state as proposed to check the validity of the scheme.  

The paper is organized as follows. In section 2, we briefly review the HHE method and the algebra of SSQM to construct the one-parameter family of isospectral potential $\hat{U}(\lambda,\rho)$. The results of our calculation are presented in section 3 while conclusions are drawn in section 4. 

\section{Theoretical Scheme}
\subsection{Hyperspherical Harmonics Expansion Method}
In the three-body cluster model of the nuclei $^A$C $\equiv ^{A-2}$C + 2n, the relatively heavy core $^{A-2}$C is labeled as particle 1, and two valence neutrons are labelled as particle 2 and 3 respectively. Thus, there are three possibile partitions for the choice of Jacobi coordinates. In any chosen partition, say in the $i^{th}$ partition, particle labelled $i$ plays the role of spectator while the remaining two paricles form the interacting pair. In this partition the Jacobi coordinates are defined as:
\begin{gather}
\left. \begin{array}{lcl}
\vec{x_{i}} &=& a_i(\vec{r_{j}} - \vec{r_{k}})\\
\vec{y_{i}} &=& \frac{1}{a_i} \left(\vec{r_{i}} -\frac{m_{j}\vec{r_{j}} + m_{k} \vec{r_{k}}}{ m_{j} + m_{k}} \right)\\
\vec{R}&=&  \frac{\sum_{i=1}^3 m_{i}\vec{r_{i}}}{M}\\
\end{array} \right]    \label{eq01}\end{gather}
where $i,j,k$ form a cyclic permutation of 1,2,3. The parameter $a_i= \left[\frac{m_{j} m_{k}M}{m_{i}(m_{j}+m_{k})^{2}} \right]^{\frac{1}{4}}$; $m_{i}, \vec{r_{i}}$ are the mass and position of the $i^{th}$ particle and $M(=\sum_{i=1}^3m_{i})$, $\vec{R}$ are the total mass and position of the centre of mass (CM) of the system. Then in terms of Jacobi coordinates, the relative motion of the three-body system can be described by the equation  \begin{gather}
\left. \begin{array}{l}
[-\frac{\hbar^{2}}{2\mu} \left(\nabla_{x_{i}}^{2}+\nabla_{y_{i}}^{2}\right) +V_{jk} (\vec{x_{i}}) +V_{ki} (\vec{x_{i}}, \vec{y_{i}}) +\\
\left.  V_{ij} (\vec{x_{i}}, \vec{y_{i}})-E \right] \Psi (\vec{x_{i}}, \vec{y_{i}}) = 0
\end{array} \right]
\label{eq02}\end{gather}
where ${\mu = \left[ \frac{m_{i} m_{j} m_{k}}{M} \right]^{\frac{1}{2}}}\rightarrow$ is the reduced mass of the system, $V_{ij}$ represents the interaction potential between the particles labelled $i$ and $j$ respectively. Hyperspherical variables are introduced through Jacobi coordinates as: $x_{i} = \rho \cos \phi_{i}$; $y_{i}= \rho \sin \phi_{i}$; $\phi_i=\tan^{-1}(\frac{y_i}{x_i})$; $\rho=\sqrt{x_i^2+y_i^2}$. Here, the hyperradius $\rho$ together with five angular variables $\Omega_{i} \rightarrow \{\phi_{i}, \theta_{x_{i}}, {\cal \phi}_{x_{i}}, \theta_{y_{i}}, {\cal \phi}_{y_{i}} \}$ constitute hyperspherical coordinates of the system. The Schr\"{o}dinger equation in hyperspherical variables $(\rho, \Omega_{i})$ becomes
\begin{gather}
\left. \begin{array}{l}
\left\{-\frac{\hbar^{2}}{2\mu}\left(\frac{1}{\rho^5} \frac{\partial^2}{\partial\rho^2}+ \frac{4}{\rho}\frac{\partial}{\partial\rho}-\frac{\hat{{\cal K}}^{2}(\Omega_{i})}{\rho^{2}}\right)+ V(\rho, \Omega_{i})-E\right\}\\
 \Psi(\rho, \Omega_{i})=0
\end{array} \right] 
\label{eq03}\end{gather}
In Eq. (\ref{eq03}) $V(\rho, \Omega_{i}) = V_{jk} + V_{ki} + V_{ij}$ is the total interaction potential in the $i^{th}$ partition and $\sl{\hat{{\cal K}}^{2}}(\Omega_{i})$ is the square of the hyperangular momentum operator satisfying the eigenvalue equation  
\begin{gather}
\hat{{\cal K}}^{2}(\Omega_{i}) {\cal Y}_{K \alpha_{i}}(\Omega_{i}) = K (K + 4) {\cal Y}_{K \alpha_{i}}(\Omega_{i})
\label{eq04} \end{gather}
$K$ is the hyperangular momentum quantum number and $\alpha_{i}$ $\equiv\{l_{x_{i}}, l_{y_{i}}, L, M \}$, ${\cal Y}_{K\alpha_{i}}(\Omega_{i})$ are the hyperspherical harmonics (HH) for which a closed analytic expressions can be found in ref. \cite{cobis-1997}. 
\newline In the HHEM, $\Psi(\rho, \Omega_{i})$ is expanded in the complete set of HH corresponding to the partition "$i$" as  
\begin{gather}
\Psi(\rho, \Omega_{i}) = \sum_{K\alpha_{i}}\frac{\psi_{K\alpha_{i}} (\rho)}{\rho^{5/2}} {\cal Y}_{K\alpha_{i}}(\Omega_{i})
\label{eq05} \end{gather} 
Use of Eq. (\ref{eq05}), in Eq. (\ref{eq03}) and application of the orthonormality of HH leads to a set of coupled differential equations (CDE) in $\rho$
\begin{gather}
\left. \begin{array}{l}
\left\{-\frac{\hbar^{2}}{2\mu}\frac{d^{2}}{d\rho^{2}}
+\frac{\hbar^{2}}{2\mu}\frac{(K+3/2)(K+5/2)}{\rho^2}-E\right\} 
\psi_{K\alpha_{i}}(\rho)\\
+\sum_{K^{\prime}\alpha_{i}^{\prime}} {\cal M}_{K\alpha_{i}}^{K^{\prime}\alpha_{i}^{\prime}}\psi_{K^{\prime}\alpha_{i}^{\prime}}(\rho)=0.
\end{array} \right]
\label{eq06} \end{gather}
where \begin{gather}
{\cal M}_{K\alpha_{i}}^{K^{\prime}\alpha_{i}^{\prime}} = \int {\cal Y}_{K\alpha_{i}}^{*}(\Omega_{i}) V(\rho, \Omega_{i}) {\cal Y}_{K^{\prime} \alpha_{i}^{~\prime}}(\Omega_{i}) d\Omega_{i}.
\label{eq07}\end{gather}
The infinite set of CDE's represented by Eq. (\ref{eq06}) is truncated to a finite set by retaining all K values up to a maximum of $K_{max}$ in the expansion (\ref{eq05}). For a given $K$, all allowed values of $\alpha_{i}$ are included. The size of the basis states is further restricted by symmetry requirements and associated conserved quantum numbers. The reduced set of CDE's are then solved by adopting hyperspherical adiabatic approximation (HAA) \cite{ballot-1982}. In HAA, the CDE's are approximated by a single differential equation assuming that the hyper radial motion is much slower compared to hyperangular motion. For this reason, the angular part is first solved for a fixed value of $\rho$. This involves diagonalization of the potential matrix (including the hyper centrifugal repulsion term) for each $\rho$-mesh point and choosing the lowest eigenvalue $U_0(\rho)$ as the lowest eigen potential \cite{das-1982}. Then the energy of the system is obtained by solving the hyper radial motion for the chosen lowest eigen potential ($U_0(\rho)$), which is the effective potential for the hyper radial motion 
\begin{gather}
\left\{-\frac{\hbar^{2}}{2\mu}\frac{d^{2}}{d\rho^{2}} + U_0(\rho) - E \right\} \psi_{0}(\rho) = 0
\label{eq08}\end{gather}
Renormalized Numerov algorithm briefed by Johnson 1978 \cite{johnson-1978}  subject to appropriate boundary 
conditions in the limit $\rho \rightarrow 0$ and $\rho\rightarrow\infty$ 
is then used to solve Eq. (\ref{eq08}) for E ($\leq E_0<0$). The hyper-partial waves $\psi_{K\alpha_{i}}(\rho)$ are given by  
\begin{gather}
\psi_{K \alpha_{i}}(\rho) = \psi_{0}(\rho) \chi_{K \alpha_{i},0}(\rho)
\label{eq09}\end{gather}
where $\chi_{K \alpha_{i},0}(\rho)$ is the ${(K\alpha_{i})}^{th}$ element of the eigenvector, corresponding to the lowest eigen potential $U_0(\rho)$.

\subsection{Construction of Isospectral Potential}
In this section we present a bird's eye view of the scheme of the construction of one parameter family of isospectral potentials. We have from Eq. (\ref{eq08}) \begin{gather}
U_0(\rho) = E_0 + \frac{\hbar^{2}}{2\mu}\frac{\psi_0^{\prime\prime}(\rho)}{\psi_0(\rho)}\label{eq10} \end{gather}
In 1-D supersymmetric quantum mechanics, one defines a superpotential for a system in terms of its ground state wave function ($\psi_{0}$) \cite{cooper-1995} as
\begin{gather}
{\cal S}(\rho)= -\frac{\hbar}{\sqrt{2m}}\frac{\psi_{0}^{\prime}(\rho)}{\psi_{0}(\rho)}.
\label{eq11}\end{gather}
The energy scale is next shifted by the ground state energy $(E_{0})$ of the 
potential $U_0(\rho)$, so that in this shifted energy scale the new potential become
\begin{gather} 
U_1(\rho) = U_0(\rho) - E_{0}=\frac{\hbar^{2}}{2\mu}\frac{\psi_0^{\prime\prime}(\rho)}{\psi_0(\rho)}\label{eq12}\end{gather} 
having its ground state at zero energy. One can then easily verify that $U_1(\rho)$ is expressible in terms of the superpotential via the Riccati equation
\begin{gather}
U_1(\rho) = {\cal S}^{2}(\rho) - \frac{\hbar}{\sqrt{2m}}{\cal S}^{\prime }(\rho).
\label{eq13}\end{gather}
By introducing the operator pairs
\begin{gather}
\left. \begin{array}{lcl}
A^{\dag}&  =&  -\frac{\hbar}{\sqrt{2m}}\frac{d}{d\rho}+{\cal S}(\rho)\\
A  &=&  \frac{\hbar}{\sqrt{2m}}\frac{d}{d\rho}+{\cal S}(\rho)
\end{array} \right\}
\label{eq14}\end{gather}
the Hamiltonian for $U_1$ becomes
\begin{gather}
H_1 = -\frac{\hbar^{2}}{2m} \frac{d^{2}}{d\rho^{2}} + U_1(\rho) = {\cal A}^{\dag}{\cal A}.
\label{eq15}
\end{gather}
The pair of operators ${\cal A}^{\dag}, {\cal A}$ serve the purpose of creation and annihilation of nodes in the wave function. Next, we introduce a partner Hamiltonian $H_{2}$, corresponding to the SUSY partner potential $U_2$ of $U_1$ as
\begin{gather}
H_{2}=-\frac{\hbar^{2}}{2m}\frac{d^{2}}{d\rho^{2}}+U_2(\rho)={\cal A}{\cal A}^{\dag}
\label{eq16}\end{gather}
where 
\begin{gather}
U_2(\rho)={\cal S}^{2}(\rho)+\frac{\hbar}{\sqrt{2m}}W^{\prime}(\rho).
\label{eq17}\end{gather}
Energy eigen values and wavefunctons corresponding to the SUSY partner Hamiltonians $H_1$ and $H_2$ are connected via the relations
\begin{gather}
\left. \begin{array}{lcl}
E_n^{(2)} & = & E_{n+1}^{(1)}, E_0^{(1)}=0 \; (n=0, 1, 2, 3,...),\\
\psi_n^{(2)}&=&\frac{1}{\sqrt{E_{n+1}^{(1)}}}A\psi_{n+1}^{(1)}\\
\psi_{n+1}^{(1)}&=&\frac{1}{\sqrt{E_n^{(2)}}}A^{\dagger}\psi_{n}^{(2)}\\
\end{array} \right\}\
\label{18} \end{gather}
where $E_{n}^{(i)}$ represents the energy of the $n^{th}$ excited state of $H_{i}$ (i=1, 2). Thus $H_{1}$ and $H_{2}$ have identical spectra, except the fact that the partner state of $H_{2}$ corresponding to the ground state of $H_{1}$ is absent in the spectrum of $H_{2}$ \cite{cooper-1995}, i.e., spectra of $H_2$ has one state less than that of $H_1$. Hence the potentials $U_1$ and $U_2$ are {\bf not strictly isospectral}. 
\newline However, one can construct, a one parameter family of {\bf strictly isospectral} potentials $\hat{U_1}(\lambda, \rho)$, explointing the fact that for a given $U_1(\rho)$, $U_0(\rho)$ and ${\cal S}(\rho)$ are not unique (see Eqs. (12) \& (13)), since the Riccati equation is a nonlinear one. Following \cite{cooper-1995} \cite{khare-1989} \cite{darboux-1882}, it can be shown that the most general superpotential satisfying Riccati equation for $U_1(\rho)$ Eq.(\ref{eq13}) is given by
\begin{gather}
\hat{S}(\rho)=W(\rho)+\frac{\hbar}{\sqrt{2m}}\frac{d}{d\rho}\log [I_{0}(\rho) +
\lambda]
\label{eq19}
\end{gather}
where $\lambda$ is a constant of integration, and $I_0$ is given by 
\begin{gather}
I_{0}(\rho)={\displaystyle\int}_{\rho^{\prime}=0}^{\rho} {[\psi_{0}(\rho^{\prime})]}^{2} d\rho^{\prime},
\label{eq20}\end{gather}
in which $\psi_0(\rho)$ is the normalized ground state wave function of $U_0(\rho)$. The potential
\begin{gather}
\left. \begin{array}{lcl}
\hat{U_1}(\lambda, \rho)&=& \hat{S}^{2}(\rho)-\frac{\hbar}{\sqrt{2m}}\hat{{\cal S}}^{\prime}(\rho) \\
&=& U_1(\rho) - 2\frac{\hbar^{2}}{2m}\frac{d^{2}}{d\rho^{2}}\log [I_0(\rho) + \lambda], 
\end{array} \right\}\
\label{eq21}
\end{gather}
has the same SUSY partner $U_2(\rho)$. $\hat{U_1}(\lambda, \rho)$ has its ground state at zero energy with the corresponding wavefunction given by  
\begin{gather}
\hat{\psi_1}(\lambda, \rho)= \frac{\psi_1}{I_0+\lambda}.
\label{eq22}\end{gather}
Hence, potentials $\hat{U_1}(\lambda, \rho)$ and $U_1(\rho)$ are {\bf strictly isospectral}. The parameter $\lambda$ is arbitrary in the intervals $-\infty<\lambda<-1$ and $0<\lambda<\infty$. $I_{0}(\rho)$ lies between 0 and 1, so the interval $-1\leq \lambda\leq 0 $ is forbidden, in order to bypass singularities in $\hat{U_1}(\lambda, \rho)$. For $\lambda \rightarrow \pm \infty$, $\hat{U_1} \rightarrow U_1$ and for $\lambda \rightarrow 0+$, $\hat{U_1}$ develops a narrow and deep attractive well near the origin. This well-barrier combination effectively traps the particle giving rise to a sharp resonance. 
For fruitful representation of the resonaces we adopt bound state in continuum (BIC) technique in which we solve Eq. (\ref{eq08}) numerically for a positive energy E subject to the boundary condition $\psi_E(0) = 0$ to get the wave function $\psi_E(r)$. This $\psi_E(r)$ is
not square integrable, and it oscillates as r increases. Following Pappademos et al 1993 \cite{pappademos-1993} one can verify by direct substitution that
\begin{gather}
\hat{\psi}_E(r,\lambda) = \frac{\psi_E(r)}{I(r)+\lambda}
\label{eq23} \end{gather}
where 
\begin{gather}
I(r) = \int_0^r |\psi_E(r^{\prime})|^2dr^{\prime}
\label{eq24}\end{gather}
satisfies Eq. (\ref{eq08}) with $U_0(r)$ replaced by 
\begin{gather}
\hat{U}(r,\lambda) = U_0(r) - \frac{\hbar^2}{2\mu}\left[\frac{4\psi_E(r)\psi_E(r)^{\prime}}{I(r)+\lambda} - \frac{2\psi_E(r)^4}{(I(r)+\lambda)^2}\right]
\label{eq25}\end{gather}
This method has been 
tested successfully for 3D finite square-well potential by Das et al 2001 \cite{das-2001} for an arbitrary particle of mass m, for Gaussian type potential to the core plus two-neucleon three-body cluster model in the case of- $^6$He by Dutta et al 2003 \cite{dutta-2003}, $^6$Li by Dutta et al 2004 \cite{dutta-2004}, $^{22}$C by Hasan et al 2019 \cite{hasan-2019}, for Wood-saxon potential in two-body model calculation of $^{15}$C by Mahapatra et al 2011 \cite{mahapatra-2011}, for density dependent M3Y potential in the case of- $^{11}$Be by Dutta et al 2014 \cite{dutta-2014}, $^{15}$Be by Dutta et al 2018 \cite{dutta-2018}.

In an attempt to  search for the correct resonance energy, we compute the probability (P) of finding the system within the potential well region of the potential $\hat{U_1}(\lambda, \rho)$ corresponding to the energy $E$ ($>0$) by integrating the norm of the wavefunctions up to the top of the barrier:
\begin{gather}
P(E)=\int_{\rho^{\prime}=0}^{\rho_B} |\hat{\psi_E}(\rho^{\prime},\lambda)|^2d\rho^{\prime}
\label{eq26}\end{gather}
in which $\rho_B$ indicates position of the top of the barrier in $\hat{U_1}(\lambda, \rho)$ for a chosen $\lambda$. Here $\hat{\psi_E}(\lambda, \rho)$ representing solution for $\hat{U_1}(\lambda, \rho)$ for a positive energy $E$, is normalized to have a constant amplitude in the asymptotic region. Plot of the quantity $P(E)$ as a function of $E$ ($E > 0$) shows a peak at the resonance energy $E=E_R$. Choice of $\lambda$ has to be made judiciously to avoid numerical errors in handling extremely narrow well for $\lambda\rightarrow 0+$. The width of resonance can be obtained from the mean life of the state using the energy-time uncertainty relation. The mean life is reciprocal to the decay constant. And the decay constant is the product of the number of hits per unit time on the barrier and the corresponding probability of tunneling out through the barrier.
\section{Results and discussions}
We apply the above scheme to the $^{18,20}$C nuclei which have in their ground state spin-parity $J^{\pi}=0^+$ and triplet isospin $T=1$. These nuclei also have a resonance state of the same $J^{\pi}$. Thus, the foregoing procedure starting from the ground state of $^{18, 20}$C will give $J^{\pi}=0^+, T=1$ resonance(s). Eq.(\ref{eq08}) is solved for the GPT n-n potential \cite{gogny-1970} and core-n SBB potential \cite{sack-1954}. The range parameter for the core-n potential $b_{cn}$ is slightly adjusted to match the observed ground state spectra. The calculated two-neutron separation energies ($S_{2n}$), their relative convergence (=$\frac{E(K_{max}+4)-E(K_{max})}{E(K_{max}+4)}$) and the RMS matter radii (R$_A$) for gradually increasing $K_{max}$ are listed in Table \ref{t01} for $^{18}$C and $^{20}$C. Although the computed results indicate a clear convergence trend with increasing $K_{max}$, it is far away from full convergence even at $K_{max}=24$. For this reason, we used an extrapolation technique successfully in atomic problems by Das et al 1994 \cite{das-1994}, Khan 2012 \cite{khan-2012} and in nuclear few-body problems by Khan and Das 2001 \cite{khan-2001}, to get the converged value of about 4.91 MeV for $^{18}$C and 3.51 MeV for $^{20}$C as shown in columns 2 and 4 of Table \ref{t02}. Partial contributions of the different partial waves corresponding to $l_x = 0, 1, 2, 3, 4$ to the two-neutron separation energies ($S_{2n}$) are presented in Table \ref{t03} for both of the systems. The computed data indicates a major contribution comes from $l_x=0$ component ($\sim 74\%$ in $^{18}$C and $\sim 96\%$ in $^{20}$C). The dependence of the two neutron separation energy on $K_{max}$ is depicted in Figure 1 for both of the nuclei $^{18}$C and $^{20}$C which indicates a saturation trend for increasing $K_{max}$. In Figure 2 we have shown the relative convergence trend in energies as a function of $K_{max}$ and it is seen that both follow a similar trend.

\begin{figure}[htb]
\begin{minipage}{12.1pc}
%\begin{minipage}{.98\columnwidth}
\centerline{\includegraphics[width=12pc,height=12pc]{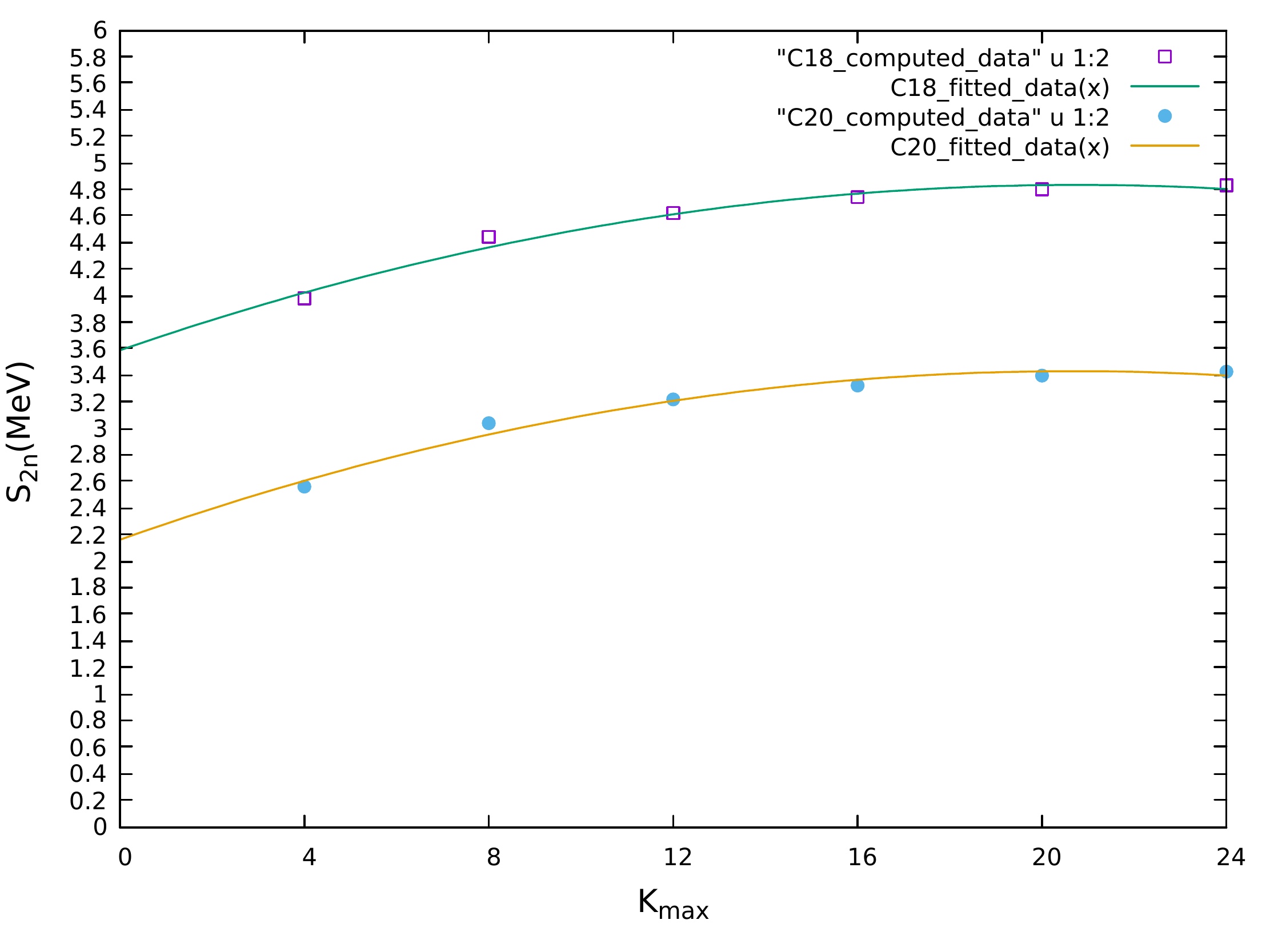}}
\caption{}{Plot of two-neutron separation energy (S$_{2n}$) as a function of K$_{max}$ for $^{18}$C (upper curve) and $^{20}$C (lower curve).}
\label{f01}
\end{minipage}\hspace{2pc}
\begin{minipage}{12.1pc}
%\begin{minipage}{.98\columnwidth}
\centerline{\includegraphics[width=12pc,height=12pc]{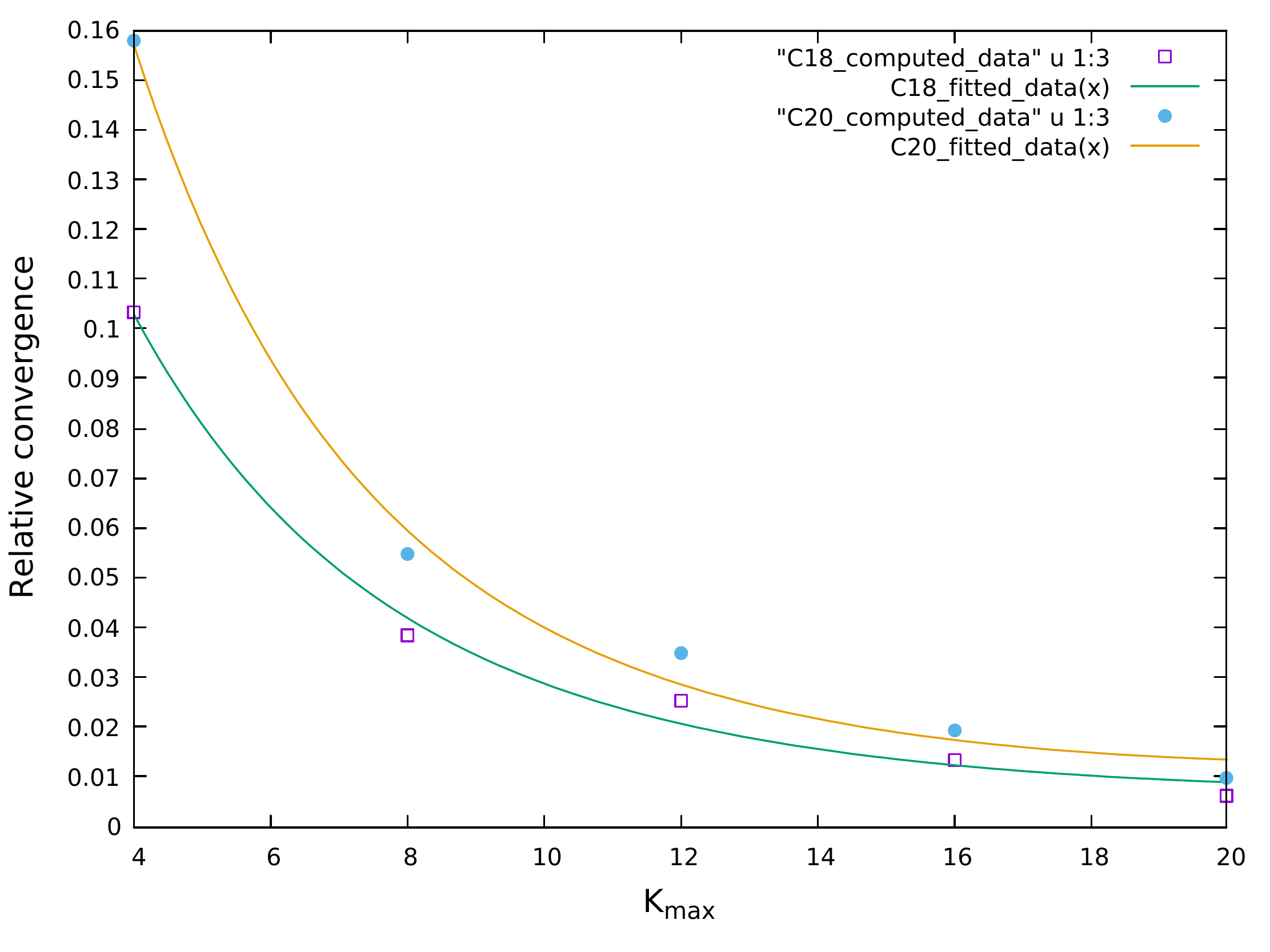}}
\caption{}{Plot of relative convergance in ground state energy with respect to increasing K$_{max}$ for $^{18}$C (lower curve) and $^{20}$C (upper curve).}
\label{f02}
%\end{minipage}{\columnwidth}
\end{minipage}
\end{figure}

In Figures 3 and 4, we have presented a 3D view of the correlation density profile of the halo nuclei $^{18}$C and $^{20}$C which indicates a halo structure with a dense core surrounded by an extended low-density tail like an envelope. 
\begin{figure}[htb]
\begin{minipage}{12pc}
%\begin{minipage}{.98\columnwidth}
\centerline{\includegraphics[width=12pc,height=12pc]{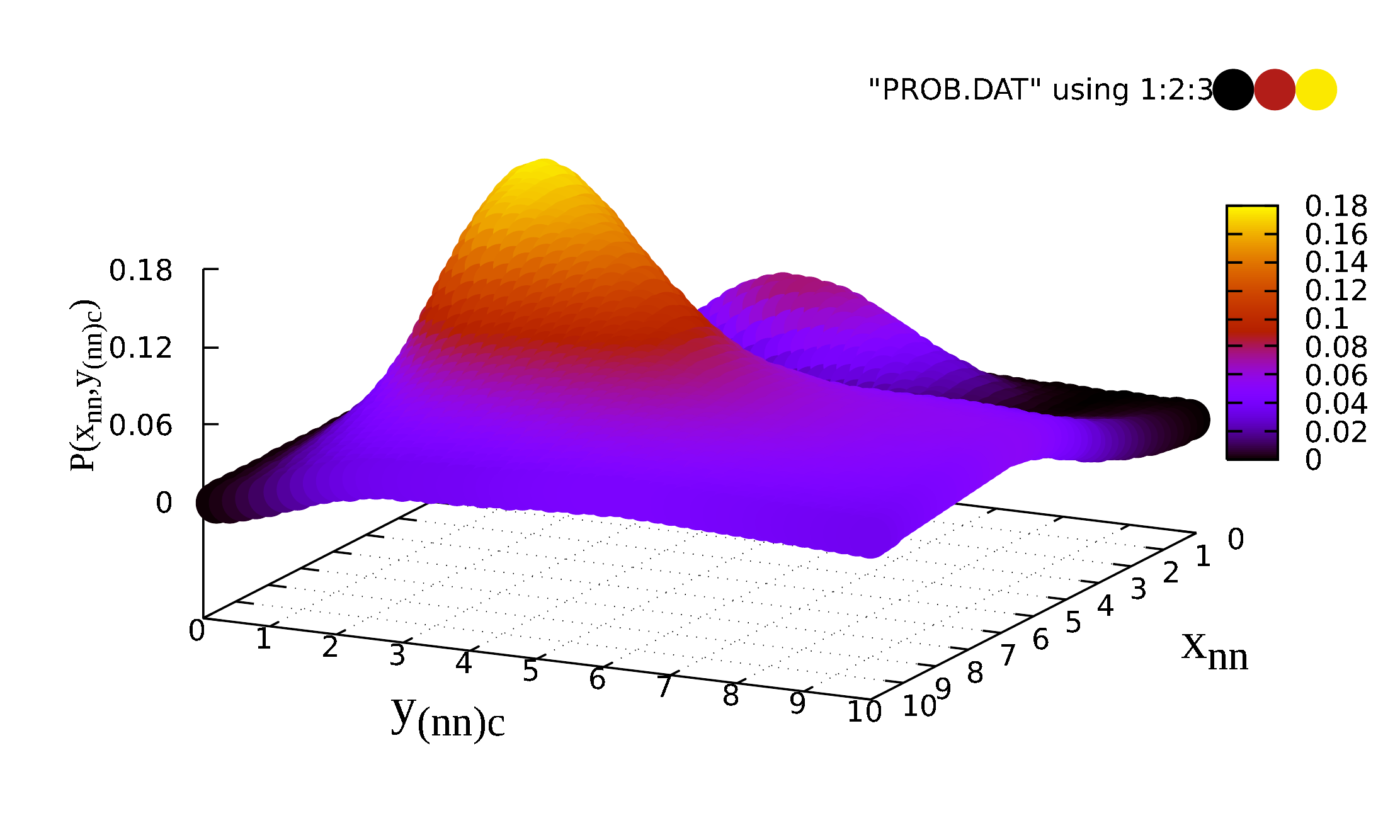}}
\caption{}{Correlation density plot for the (0$^{+}$) ground state of $^{18}$C as function of the Jacobi coordiantes $x_{nn}$ and $y_{(nn)c}$.}
\label{f03}
\end{minipage}\hspace{2pc}
\begin{minipage}{12pc}
%\begin{minipage}{.98\columnwidth}
\centerline{\includegraphics[width=12pc,height=12pc]{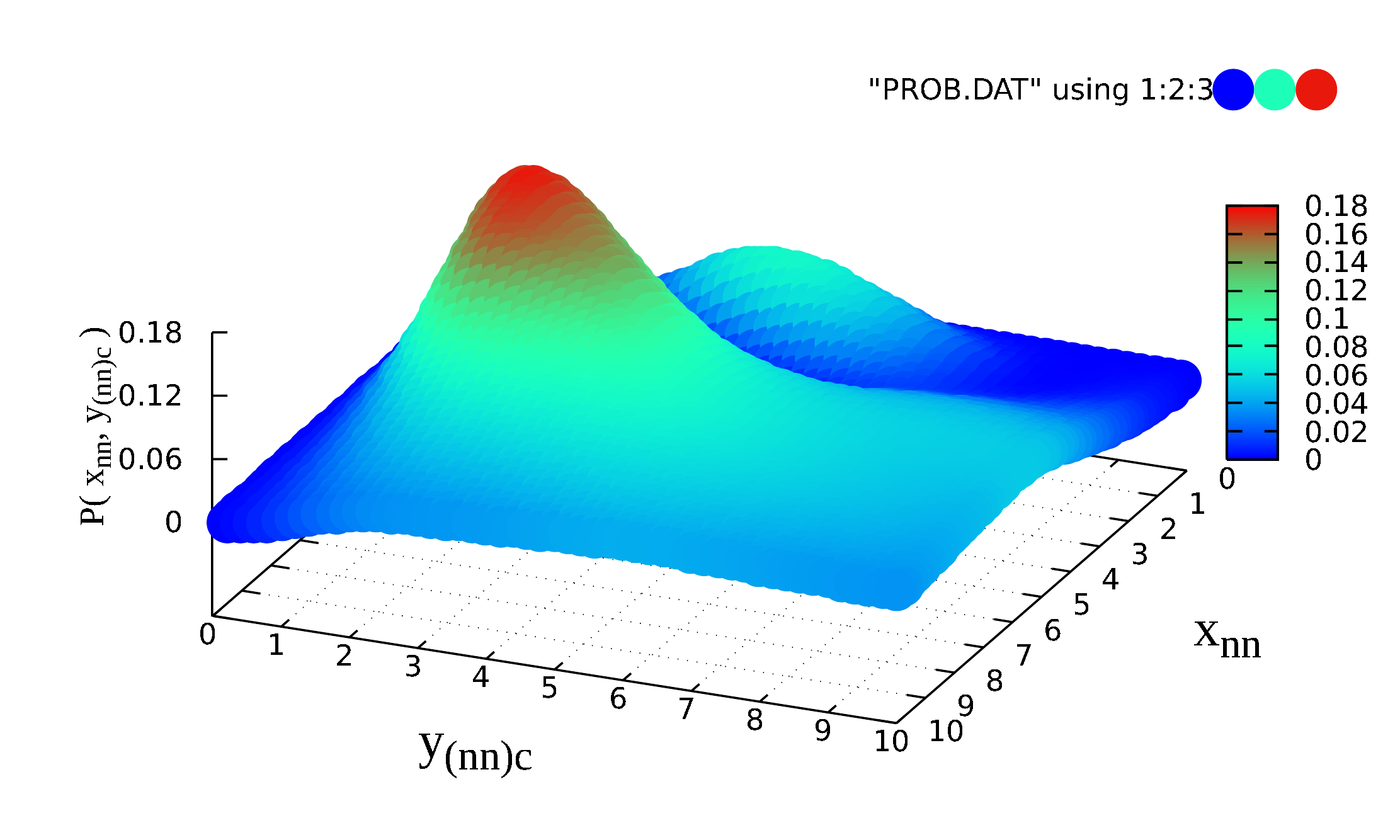}}
\caption{}{Correlation density plot for the (0$^{+}$) ground state of $^{20}$C as function of the Jacobi coordiantes $x_{nn}$ and $y_{(nn)c}$.}
\label{f04}
%\end{minipage}{\columnwidth}
\end{minipage}
\end{figure}
Similar nature is also reflected in Figures 5 and 6 where we have shown the 2D projection of the 3D probability density distribution. 
\begin{figure}[htb]
\begin{minipage}{12pc}
%\begin{minipage}{.98\columnwidth}
\centerline{\includegraphics[width=12pc,height=12pc]{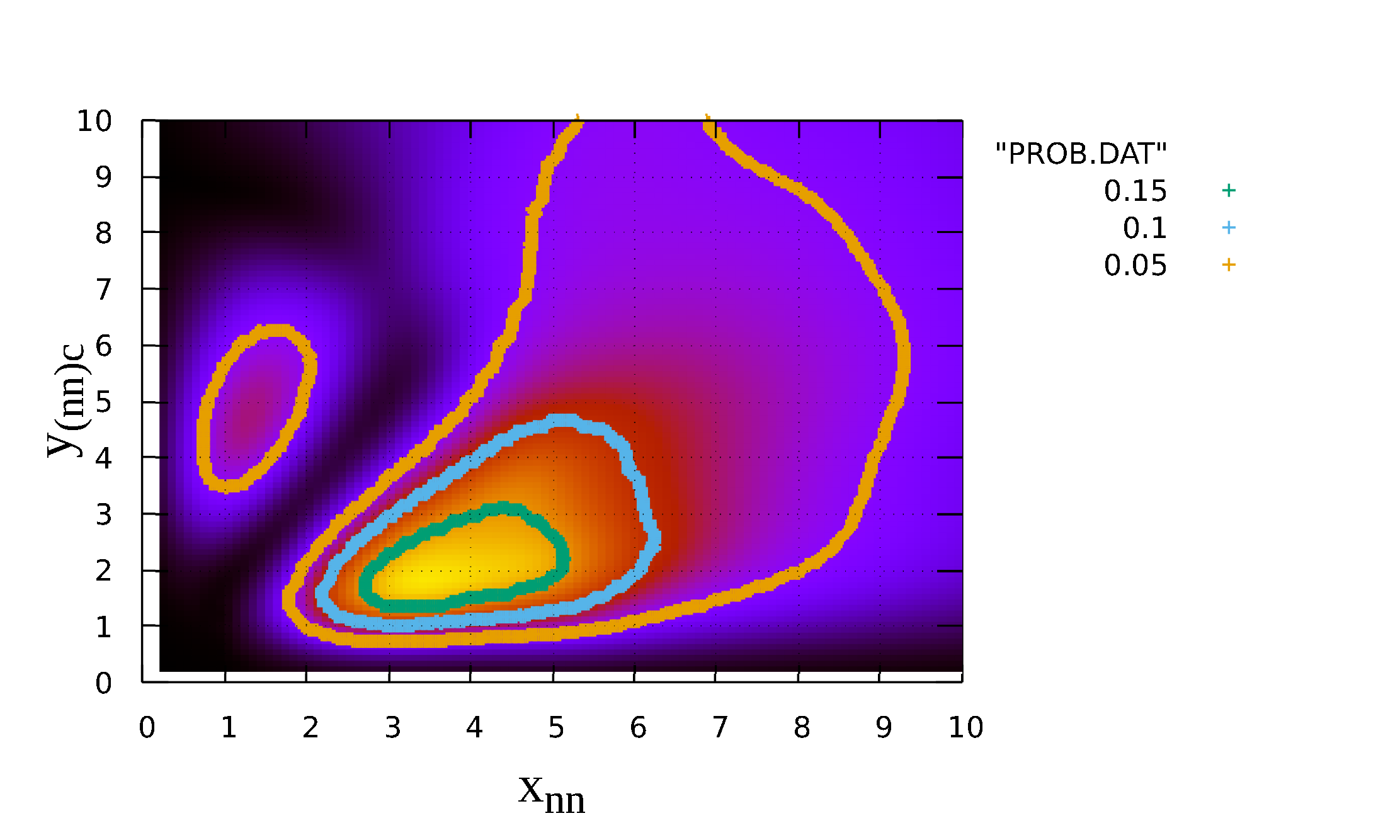}}
\caption{}{2D projection of the correlation density plot for the 0$^{+}$ state of $^{18}$C as a function of the Jacobi coordiantes $x_{nn}$ and $y_{(nn)c}$.}
\label{f05}
\end{minipage}\hspace{2pc}
\begin{minipage}{12pc}
%\begin{minipage}{.98\columnwidth}
\centerline{\includegraphics[width=12pc,height=12pc]{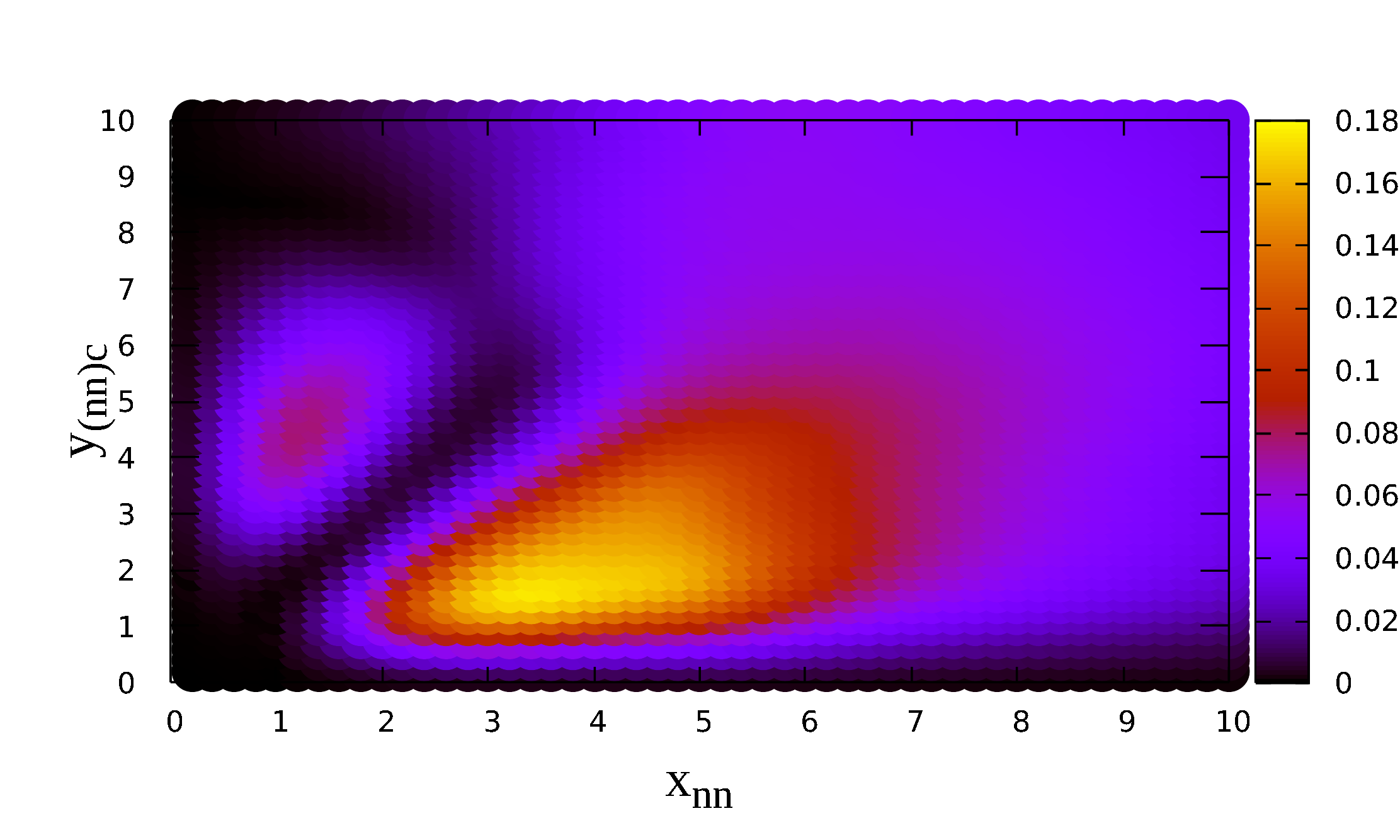}}
\caption{}{2D projection of the correlation density plot for the 0$^{+}$ state of $^{20}$C as a function of the Jacobi coordiantes $x_{nn}$ and $y_{(nn)c}$.}
\label{f06}
\end{minipage}
\end{figure}

After getting the ground state energy and wavefunctions we constructed the isospectral potential invoking principles of SSQM to investigate the resonant states. The lowest eigen potential obtained for the ground states of $^{18}$C and $^{20}$C as shown in Figures 7 and 8 by thick violet lines, exhibit shallow well followed by a wide-skinny barrier.
\begin{figure}[htb]
\begin{minipage}{12pc}
%\begin{minipage}{.98\columnwidth}
%\centering
\centerline{\includegraphics[width=12pc,height=12pc]{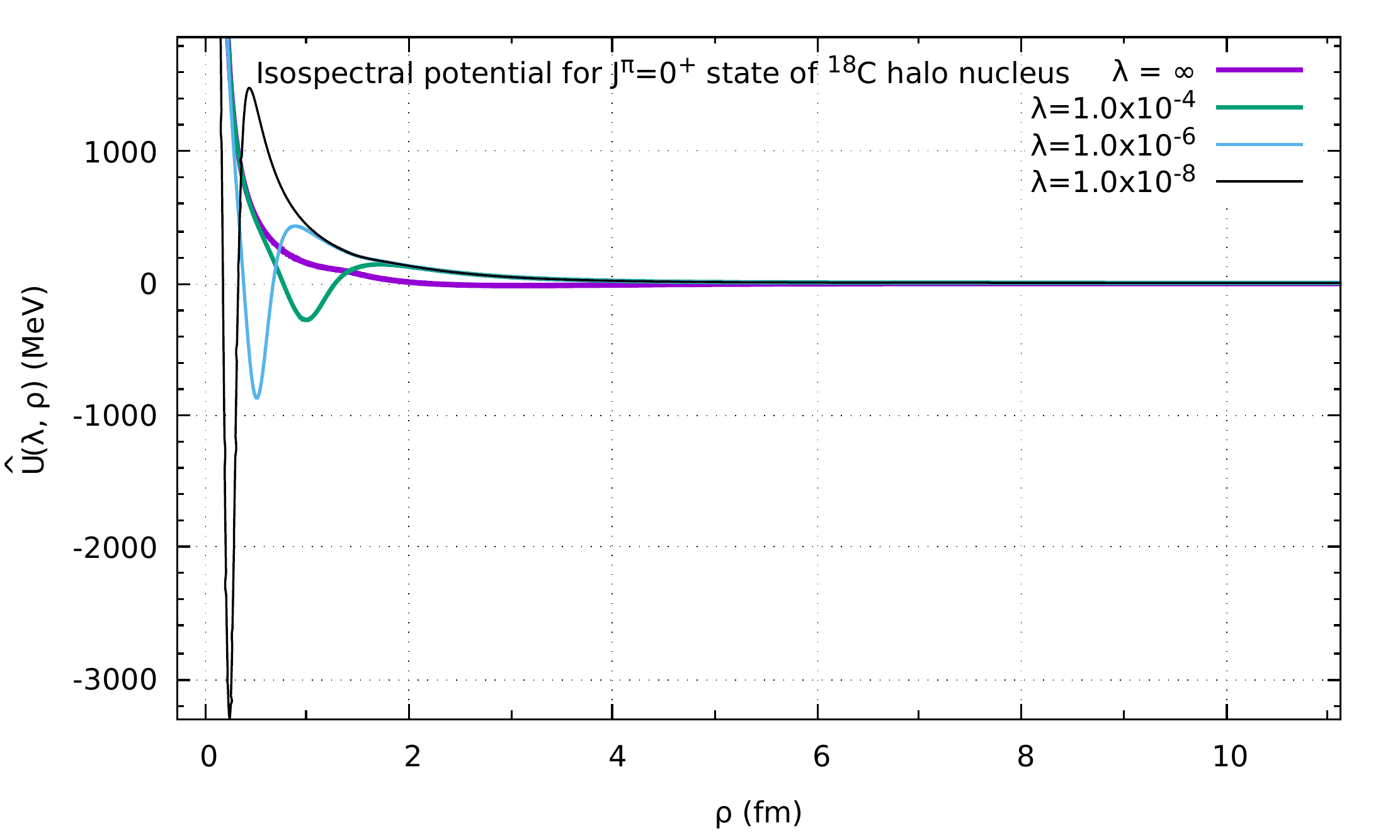}}
\caption{}{Plot of the one-parameter family of isospectral potential constructed for $0^+$ resonance in $^{18}$C for some representative values of the parameter $\lambda$ ($=\infty$(original shallow potential), $10^{-4}, 10^{-6}$ and $10^{-8}$.}
\label{f07}
\end{minipage}\hspace{2pc}
\begin{minipage}{12pc}
%\begin{minipage}{.98\columnwidth}
\centerline{\includegraphics[width=12pc,height=12pc]{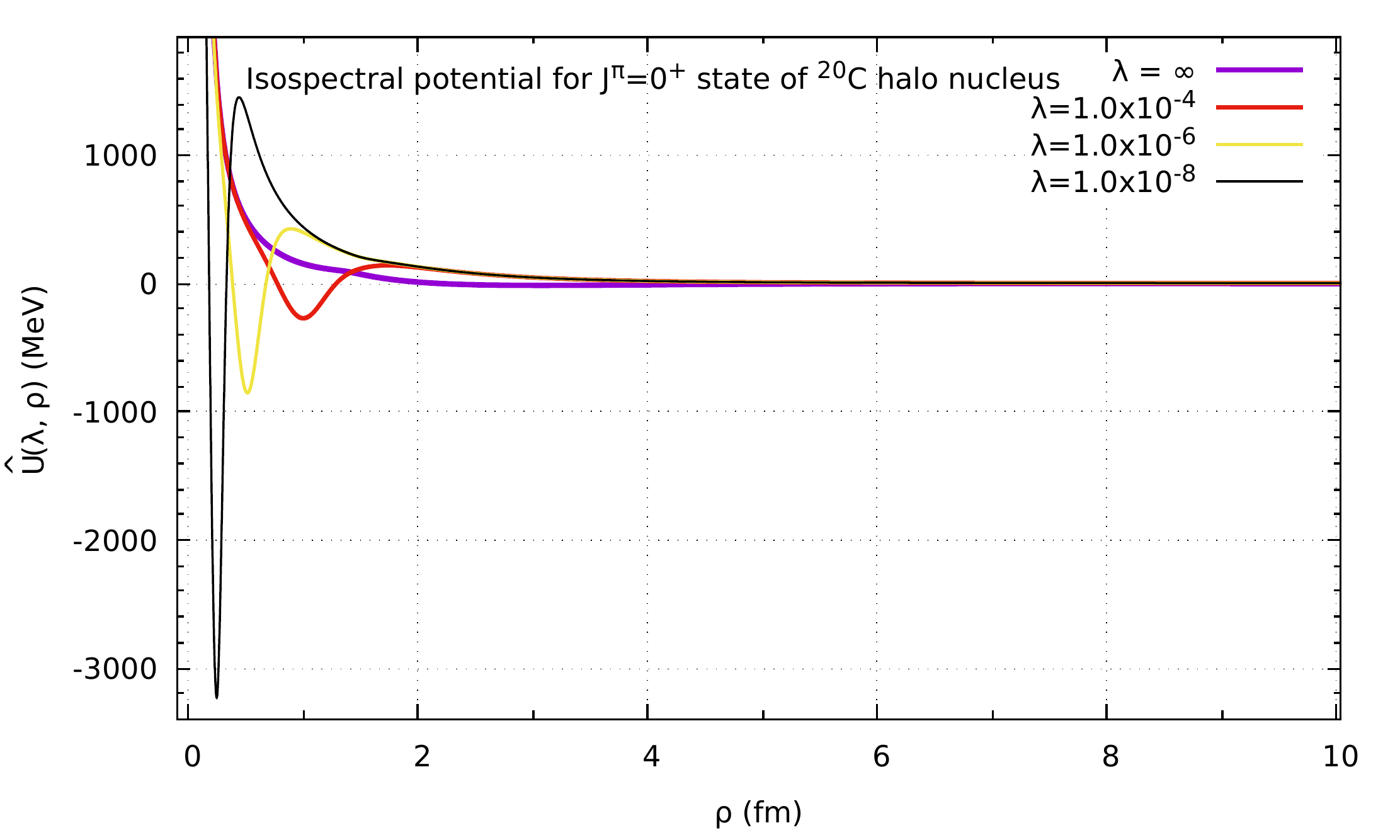}}
\caption{}{Plot of the one-parameter family of isospectral potential constructed for $0^+$ resonance in $^{20}$C for some representative values of the parameter $\lambda$ ($=\infty$(original shallow potential), $10^{-4}, 10^{-6}$ and $10^{-8}$.}
\label{f08}
\end{minipage}
\end{figure}
%.....

This kind of shallow-well and skinny barrier combination supports weak bound states with a large spatial extension of the wavefunction shown by thick black lines in Figures 9 \& 10. 
\begin{figure}[htb]
\begin{minipage}{12pc}
%\begin{minipage}{.98\columnwidth}
%\centering
\centerline{\includegraphics[width=12pc,height=12pc]{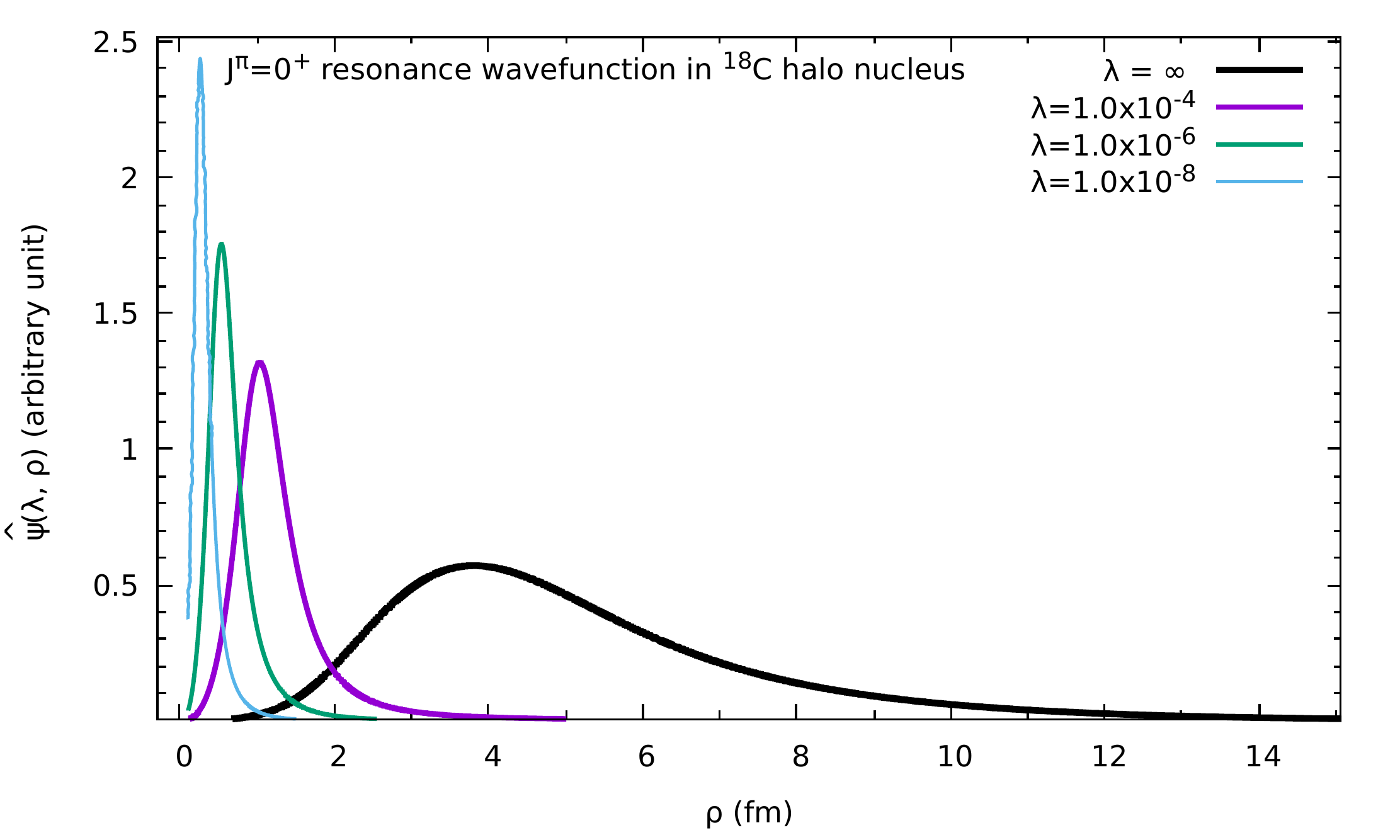}}
\caption{}{Plot of the wavefunctions computed at energy of $0^+$ resonance in $^{18}C$ for some representative values of the parameter $\lambda$ ($=\infty$(correspond to original shallow potential), $10^{-4}, 10^{-6}$ and $10^{-8}$.}
\label{f09}
\end{minipage}\hspace{2pc}
\begin{minipage}{12pc}
%\begin{minipage}{.98\columnwidth}
\centerline{\includegraphics[width=12pc,height=12pc]{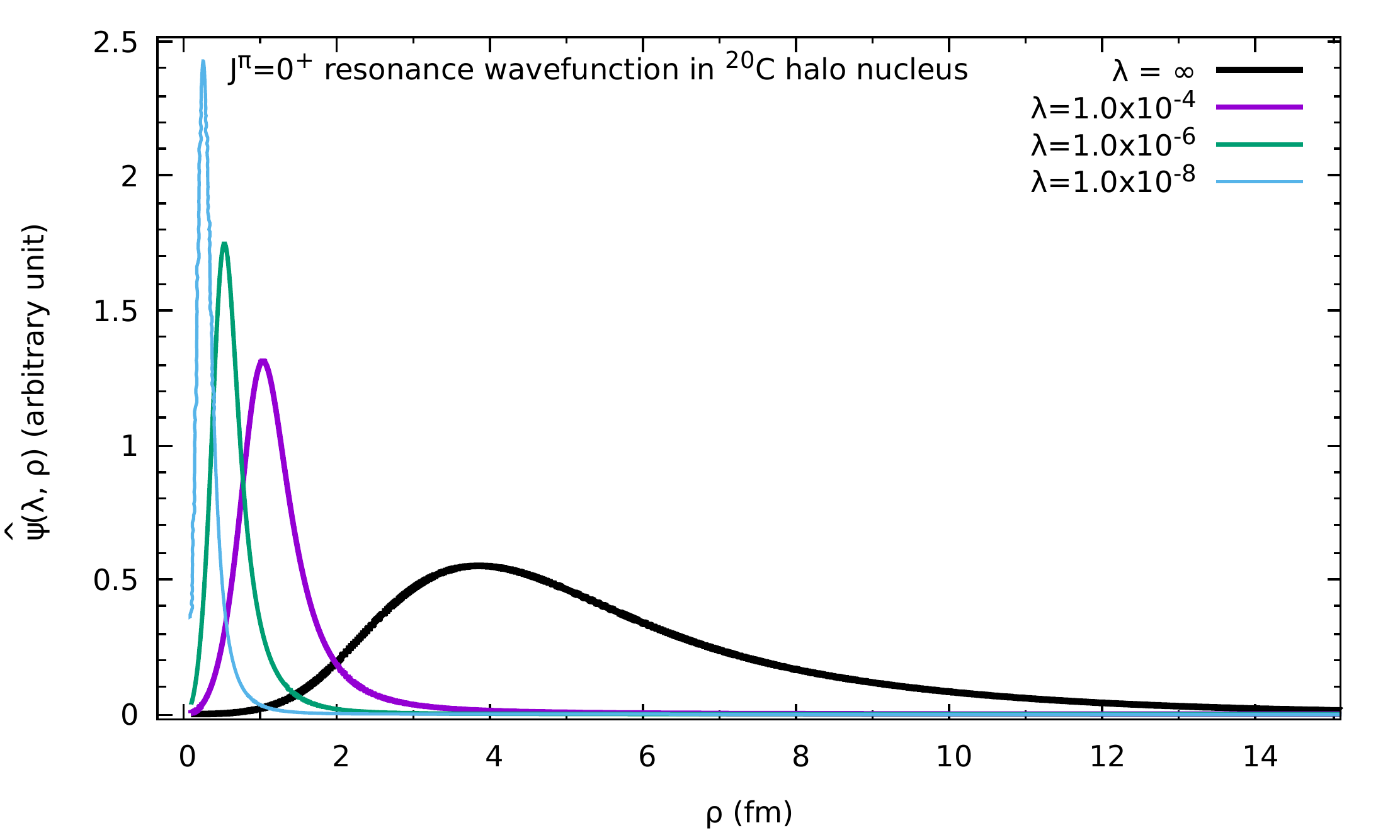}}
\caption{}{Plot of the wavefunctions computed at energy of $0^+$ resonance in $^{20}C$ for some representative values of the parameter $\lambda$ ($=\infty$(correspond to original shallow potential), $10^{-4}, 10^{-6}$ and $10^{-8}$.}
\label{f10}
\end{minipage}
\end{figure}
%.....

Such potentials may also support broad resonances for the far extended wavefunctions. Accurate computation of resonance energy and width of such kinds of broad resonances is a challenging numerical task due to large computational errors. Hence, we constructed the one-parameter family of isospectral potentials $\hat{U}(\lambda, \rho)$ following Eq.(\ref{eq21}) by appropriate selection of $\lambda$ values to ensure narrow and sufficiently deep potential well following a high barrier. The constructed potentials are shown in Figures 7 and 8 for some representative values of $\lambda$ for both of the systems considered here. The enhanced well-barrier combination effectively traps the particles to form a strong resonant state. With gradual decrease in the $\lambda$ values from $\infty$ towards 0+ enhances both the depth of the well and height of the barrier. It is also observed that the overall width of the well-barrier combination decreases with a decrease in the value of $\lambda$ together with the shifting of the positions of minima of the well and maxima of the barrier towards the origin.  Calculated parameters of the isospectral potential for some representative $\lambda$ values in [$+\infty,0$] are presented in Table \ref{t04}, in which $\lambda=\infty$ corresponds to the original effective potential $U_0(\rho)$ of Eq. (\ref{eq08}) as can be verified from Eqs. (\ref{eq21}) with a reference to Eq. (\ref{eq12}). One can, for example, note from columns under $^{18}$C in Table \ref{t04} that, when $\lambda$ decreases from 0.1 to 0.0001, the depth of the potential well increases from -24.2 MeV at 2.7 fm to -247.6 MeV at 1.3 fm while the height of the barrier increases from 5.4 MeV at 5.1 fm to 121.5 MeV at 1.9 fm respectively. The same trend can also be noted for $^{20}$C. Thus the application of SSQM produces a {\bf dramatic effect} in the isospectral potential $\hat{U_1}(\lambda, \rho)$ as $\lambda$ approaches 0+. Still smaller positive values of $\lambda$ are not desirable since that will make the potential well too narrow to compute the wave functions accurately by a standard numerical technique. 
The probability of trapping (P(E)) of the particle within the enhanced well-barrier combination as a function of energy E displayed as Figures 11 and 12 exhibit resonance peak at the energy $E_R\simeq 1.98$ MeV in $^{18}$C and energy $E_R\simeq 3.75$ MeV in $^{20}$C respectively. 

\begin{figure}[htb]
\begin{minipage}{12pc}
%\begin{minipage}{.98\columnwidth}
\centerline{\includegraphics[width=12pc,height=12pc]{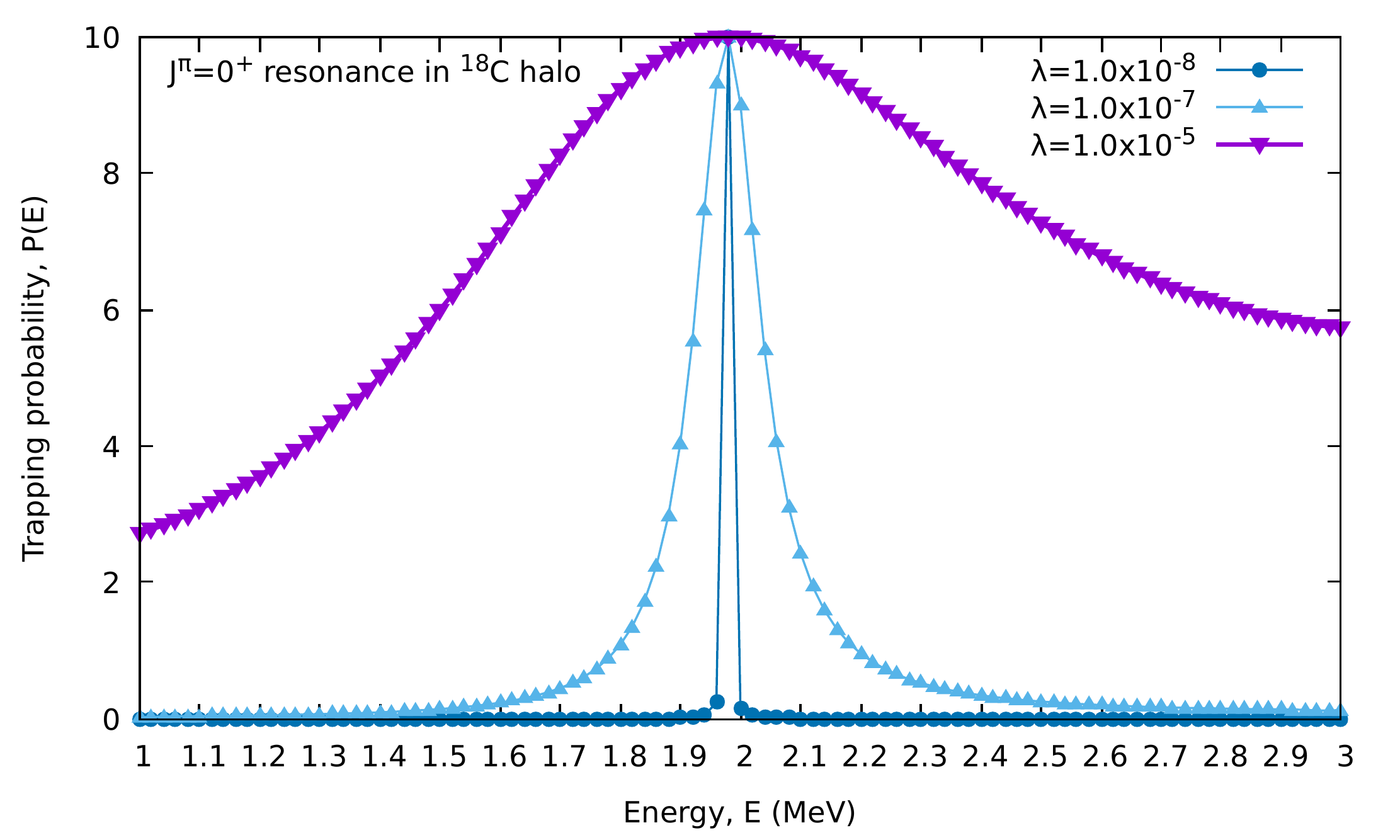}}
\caption{}{Plot of trapping probability $P(E)$ as a function of energy demostrating the $0^+$ resonant state in $^{18}C$.}
\label{f11}
\end{minipage}\hspace{2pc}
\begin{minipage}{12pc}
%\begin{minipage}{.98\columnwidth}
\centerline{\includegraphics[width=12pc,height=12pc]{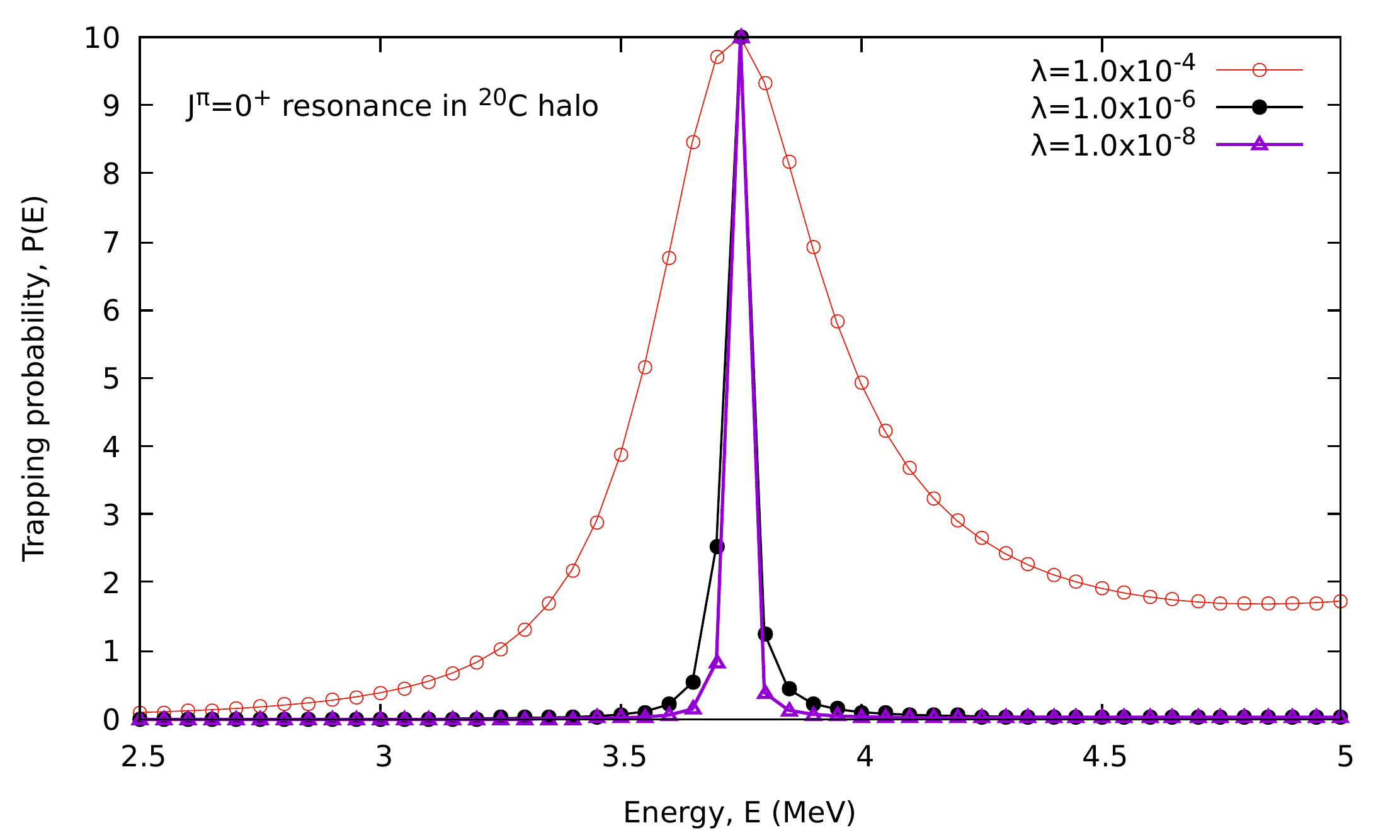}}
\caption{}{Plot of trapping probability $P(E)$ as a function of energy demostrating the $0^+$ resonant state in $^{20}C$.}
\label{f12}
\end{minipage}
\end{figure}
%%%%
%%%%
%
It is interesting to note that the resonance energy is independent of the $\lambda$ parameter. The enhancement of accuracy in the determination of $E_R$ is the principal advantage of using Supersymmetric formalism. Since $\hat{U_1}(\lambda;\rho)$ is strictly isospectral with $U(\rho)$, any value of $\lambda$ is admissible in principle. However, a judicious choice of $\lambda$ is necessary for the accurate determination of the resonance energy and to avoid delta-function type behavior of the isospectral potential. The calculated two-neutron separation energies as presented in Table \ref{t05} are in excellent agreement with the observed values of $4.910\pm 0.030$ MeV for $^{18}$C and $3.510\pm 0.240$ MeV for $^{20}$C \cite{audi1-2003} and also with results of Yamaguchi et al \cite{yamaguchi-2011}. The calculated RMS matter radii also agree fairly with the experimentally observed values found in \cite{ozawa1-2001}. A similar type of calculation adopting core-neutron two-body model for $^{15}$Be has recently been reported by Dutta et al 2018 \cite{dutta-2018}. 

%\begin{center}
\begin{table}[htbp]
\caption[]{Calculated two-neutron separation energies, their relative convergence and RMS matter radius of $^{18}$C and $^{20}$C for different $K_{max }$ in their ground states.}\small\smallskip
%\begin{tabular}{|c|c|c|c|c|c|c|}\hline\hline
\tabcolsep=3.6pt
\begin{tabular}{@{}ccccccc@{}}
\hline\hline
\
$System$ &\multicolumn{3}{c}{$^{18}$C ($^{16}$C+n+n)}&\multicolumn{3}{c}{$^{20}$C ($^{18}$C+n+n)}\\
%\cline{2-4}\cline{5-7}
\hline
&&&&&&\\[-8pt]
$K_{max}$&$S_{2n}$(MeV)&Relative Conv.&$R_{A}$&$S_{2n}$(MeV)&Relative Conv.&$R_{A}$\\\hline
&&&&&&\\[-8pt]
4 &3.97843&0.10341&2.8297&2.55712&0.15771&2.9832\\
8 &4.43727&0.03861&2.7913&3.03591&0.05480&2.9431\\
12&4.61546&0.02521&2.7647&3.21193&0.03493&2.9145\\
16&4.73481&0.01344&2.7425&3.32818&0.01944&2.8867\\
20&4.79933&0.00638&2.7233&3.39416&0.00984&2.8628\\
24&4.83013&       &2.7156&3.42789&       &2.8479\\

\hline\hline
\end{tabular}
\label{t01}
\end{table}
%\end{center}
%%%%%
%%%%%
%\begin{center}
\begin{table}[htbp]
\caption[]{Extrapolated values of two-neutron separation energies $(S_{2n}$ and their relative convergences for the ground state (J$^{\pi}$=0$^+$) of $^{18}$C and $^{20}$C.}\small\smallskip
%\centering
\tabcolsep=3.6pt
\begin{tabular}{@{}ccccc@{}}
\hline\hline
$System$&\multicolumn{2}{c}{$^{18}$C ($^{16}$C+n+n)}&\multicolumn{2}{c}{$^{20}$C ($^{18}$C+n+n)}\\
\hline
&&&&\\[-8pt]
%\cline{1-3}\cline{4-5}
$K_{max}$&$S_{2n}(MeV)$&Relative Convergence&$S_{2n}(MeV)$&Relative Convergence\\\hline
24 &4.83013120&0.00456309&3.42789570&0.00669549\\
28 &4.85227259&0.00303493&3.45100186&0.00446219\\
32 &4.86704371&0.00209706&3.46646992&0.00308899\\
36 &4.87727164&0.00149547&3.47721099&0.00220653\\
40 &4.88457636&0.00109515&3.48490054&0.00161829\\
44 &4.88993151&0.00082034&3.49054929&0.00121388\\
48 &4.89394622&0.00062664&3.49479156&0.00092838\\
52 &4.89701487&0.00048691&3.49803908&0.00072217\\
56 &4.89940043&0.00038406&3.50056708&0.00057019\\
60 &4.90128282&0.00030700&3.50256423&0.00045621\\
64 &4.90278798&0.00024834&3.50416285&0.00036934\\
68 &4.90400585&0.00020305&3.50545757&0.00030221\\
72 &4.90500182&0.00016763&3.50651729&0.00024967\\
76 &4.90582420&0.00013961&3.50739299&0.00020807\\
80 &4.90650922&0.00011721&3.50812293&0.00017479\\
84 &4.90708439&0.00009913&3.50873623&0.00014791\\
88 &4.90757089&0.00008441&3.50925527&0.00012599\\
92 &4.90798515&0.00007231&3.50969749&0.00010800\\
96 &4.90834009&0.00006232&3.51007659&0.00009311\\
100&4.90864598&0.00005399&3.51040343&0.00008069\\
104&4.90891099&0.00004693&3.51068673&0.00007028\\
108&4.90914137&0.00004119&3.51093349&0.00006149\\
112&4.90934358&0.00003612&3.51114943&0.00005405\\
116&4.90952089&0.00003186&3.51133920&0.00004769\\
120&4.90967732&0.00002821&3.51150666&0.00004224\\
124&4.90981582&0.00002507&3.51165499&0.00003755\\
128&4.90993891&0.00002235&3.51178684&0.00003349\\
132&4.91004866&0.00001999&3.51190444&0.00002996\\
136&4.91014683&0.00001794&3.51200967&0.00002689\\
140&4.91023492&0.00001614&3.51210410&0.00002419\\
144&4.91031418&0.00001456&3.51218909&0.00002184\\
148&4.91038569&0.00001318&3.51226579&0.00001976\\
152&4.91045039&0.00001195&3.51233519&0.00001792\\
156&4.91050905&0.00001086&3.51239814&0.00001629\\
160&4.91056237&0.00000989&3.51245537&0.00001484\\
164&4.91061094&0.00000903&3.51250750&0.00001355\\
168&4.91065528&0.00000826&3.51255511&0.00001239\\
172&4.91069584&0.00000757&3.51259866&0.00001137\\
176&4.91073301&0.00000695&3.51263858&0.00001044\\
180&4.91076715&0.00000639&3.51267525&0.00000960\\
184&4.91079855&0.00000589&3.51270898&0.00000885\\

........&........&........&........&........\\
$\infty$&4.91124921&  &3.51319416&\\

\hline\hline
\end{tabular}
\label{t02}
\end{table}
%\end{center}
%%%
%\begin{center}
\begin{table}[htbp]
\caption[]{Partial contribution of different $l_x$ partial waves to the two-neutron separation energies in the ground state of $^{18}$C and $^{20}$C for graduallt increasing $K_{max }$.}\small\smallskip
%\centering
\tabcolsep=3.6pt
\begin{tabular}{@{}ccccccccccc@{}}
\hline\hline
$System$ &\multicolumn{5}{c}{$^{18}$C} &\multicolumn{5}{c}{$^{20}$C}\\
%\cline{1-6}\cline{7-11}
$K_{max}$&\multicolumn{5}{c}{$E_{lx}$for $l_x$=}&\multicolumn{5}{c}{$E_{l_x}$ for $l_x$=}\\
%\cline{2-11}
\hline
&&&&&&&&&&\\[-8pt]
  &0       &1      &2      &3      &4       &0 &1&2&3&4\\
4 &2.886 &0.068&1.149&0.000&0.000&2.471&0.235&0.036&0.000&0.000\\
8 &3.283 &0.072&1.158&0.001&0.100&2.902&0.232&0.019&1.034&0.001\\
12&3.414 &0.078&1.158&0.001&0.098&3.073&0.231&0.014&1.036&0.001\\
16&3.487 &0.079&1.166&0.001&0.098&3.195&0.228&0.014&1.040&0.001\\
20&3.536 &0.081&1.181&0.001&0.096&3.262&0.225&0.013&1.049&0.001\\
24&3.566 &0.082&1.199&0.001&0.094&3.295&0.225&0.013&1.080&0.001\\
\hline\hline
\end{tabular}
\label{t03}
\end{table}
%\end{center}
%%%%%%%%%%%%%

%\begin{center}
\begin{table}[htbp]
\caption[]{Data to show the effect of the parameter $\lambda$ on the depth ($V_0$) of the well and height ($V_B$) of the barrier in the isospectral potential constructed using Eq. (\ref{eq21}) for $^{18}$C and $^{20}$C.}\small\smallskip
\tabcolsep=3.6pt
\begin{tabular}{@{}lcccccccc@{}}
\hline\hline
&&&&&&&&\\[-8pt]
$System$ &\multicolumn{4}{c}{$^{18}$C} &\multicolumn{4}{c}{$^{20}$C}\\
\cline{1-5}\cline{6-9}
$\lambda $&\multicolumn{2}{c}{Potential	Well}	&\multicolumn{2}{c}{Potential	Barrier}	&\multicolumn{2}{c}{Potential	Well}	&\multicolumn{2}{c}{Potential	Barrier}	\\
\cline{2-9}
&$V_0$&$r_0$&$V_B$&$r_B$&$V_0$&$r_0$&$V_B$&$r_B$\\
&$(MeV)$&$(fm)$&$(MeV)$&$(fm)$&$(MeV)$&$(fm)$&$(MeV)$&$(fm)$\\
\hline
&&&&&&&&\\[-8pt]
$10^5$ &-9.301  &3.092&2.702  &20.099&-11.076&3.069&3.085&7.520\\
$10^2$    &-9.311  &3.092&2.712  &20.099&-11.086&3.069&3.094&7.514\\
50     &-9.337  &3.091&2.712  &20.099&-11.163&3.065&3.106&7.504\\
1      &-11.590 &3.019&2.719  &20.095&-17.294&2.824&3.938&6.708\\
0.1    &-24.220 &2.664&5.394  &5.088&-41.931&2.320&10.924&3.983\\
$10^{-2}$   &-62.846 &2.109&19.797 &3.424&-95.444&1.811&36.751&2.840\\
$10^{-3}$  &-136.144&1.637&56.596 &2.521&-154.453&1.389&84.198&2.173\\
$10^{-4}$ &-247.602&1.274&121.482&1.937&-270.962&1.000&141.518&1.731\\
$10^{-5}$&-384.056&0.986&218.025&1.520&-479.005&0.724&241.024&1.218\\\hline\hline
\end{tabular}
\label{t04}
\end{table}
%\end{center}
%%%
%%%
\begin{table}[htbp]
%\begin{center}
\caption[]{Comparison of calculated data with those found in the literature for the two-neutron halo nuclei $^{18}$C and $^{20}$C.}\small\smallskip
\begin{tabular}{@{}lllll@{}}
\hline\hline
Nuclei&State ($J^{\pi}$)& Observables& Present work & Others work\\
\hline
&&&&\\[-8pt]
% % % %
$^{18}$C&$0^{+}$             & $S_{2n}$ (MeV)      &4.9064 &4.910 $\pm$0.030 \cite{audi1-2003}\\
&&&&4.91\cite{yamaguchi-2011}\\
&             & $R_A$ (fm)      & 2.7156 &2.82$\pm 0.04$ \cite{ozawa1-2001}\\%\cline{0-0}
&0$^{+}_{1}$&$E_{R}$ (MeV) &1.89 &-\\\hline  
$^{20}$C&$0^{+}$             & $S_{2n}$ (MeV)      &3.5065 &3.510 $\pm$0.240 \cite{audi1-2003}\\
&&&&3.51 \cite{yamaguchi-2011}\\
&             & $R_A$ (fm)      &2.8479 &2.98$\pm0.05$ \cite{ozawa1-2001}\\%\cline{2-6}
&0$^{+}_{1}$&$E_{R}$ (MeV) &3.735 &-\\
\hline\hline
\end{tabular}
\label{t05}
%\end{center}
\end{table}
%
%\newpage

\section{Summary and conclusions}
In this communication, we have investigated the ground state structure of $^{18, 20}$C using hyperspherical harmonics expansion method assuming $^{16,18}$C$+n+n$ three-body cluster model. Standard GPT \cite{gogny-1970} potential is chosen for the $n-n$ pair while a three-term Gaussian SBB potential \cite{sack-1954} with spin-orbit coupling term is used for the core-nucleon subsystems. The ground state energy and wavefunctions are obtained by numerical solution of the three-body Schr\"{o}diger equation using renormalized Numerov method algorithm \cite{johnson-1978}. The three-body effective potential $U_0(\rho)$ in Eq. (\ref{eq08}) exhibits a shallow well following a spatially extended skinny barrier (shown in Figures 7 \& 8 as thick blue lines corresponding to $\lambda =\infty$). These kinds of shallow well-barrier combinations support very weak and much broader resonances for far extended wavefunctions resulting in an erroneous numerical computation. Hence, for accurate computation of such broad resonances, a large number of partial waves need to be included in the expansion basis. But the spatially extended partial waves cause slowing down of the convergence rate and introduces more errors in the result. Hence, a one-parameter family of isospectral potential is constructed using the ground state energy and wavefunctions following the algebra of SSQM. This newly constructed one-parameter family of isospectral potential for an appropriate choice of $\lambda$ develops a narrow and deep narrow well shifted towards the origin following relatively higher and less extended barrier (as shown in Figures 7 \& 8 as lines of gradually decreasing thickness for $\lambda = \infty, 10^{-4}, 10^{-6}$, and $10^{-8}$). This enhanced well-barrier combination correctly represents the wavefunctions (as shown in Figures 9 \& 10) and resonances at the same energy as in the original shallow potential (as shown in Figures 11 \& 12) for $\lambda = 10^{-4}, 10^{-6}$, and $10^{-8}$ respectively. As summarized in Table \ref{t05} the computed observables are in excellent agreement with experimental data available in the literature. In conclusion, we may say that the scheme described here is a robust one and can be applied to any weakly bound few-body system even if the system lacks any bound state. The above facts and findings apart from validating our theoretical scheme also accomplished our goals set for this work.
%%%%
%%%%

\section*{Acknowledgements} 
This work has been supported by the computational facility at Aliah University, India.\

{\bf Conflict of interest statement:} On behalf of all authors, the corresponding author states that there is no conflict of interest.

\end{document}